\documentclass[a4paper,fleqn]{cas-dc}

\usepackage[numbers]{natbib}
\usepackage{graphicx}
\usepackage[commandnameprefix=always]{changes}
\definechangesauthor[name={Anna}, color=blue]{AF}
\usepackage{makecell} 
\usepackage{caption} 

\def\tsc#1{\csdef{#1}{\textsc{\lowercase{#1}}\xspace}}
\tsc{WGM}
\tsc{QE}

\begin{document}
\let\WriteBookmarks\relax
\def\floatpagepagefraction{1}
\def\textpagefraction{.001}

\shorttitle{}    

\shortauthors{}  

\title [mode = title]{Development and Performance of an Instrumentation Laboratory for Infrared Medical Imaging}  



%

\author[1]{Anna Frixou}[orcid=0009-0009-2658-8002]



\ead{a.frixou@cyi.ac.cy}



\affiliation[1]{organization={The Cyprus Institute},
            addressline={20 Kavafi Street}, 
            city={Nicosia},
            postcode={2121}, 
            country={Cyprus}}

\author[1,2]{Efstathios Stiliaris}[orcid=0000-0001-9496-7762]


\ead{stiliaris@phys.uoa.gr}



\affiliation[2]{organization={
Department of Physics, National and Kapodistrian University of Athens},
            addressline={Zografou, GR-157 84}, 
            city={Athens},
            country={Greece}}

\author[1,2,3]{Costas N. Papanicolas}[orcid=0000-0001-5320-0768]

\cormark[1]
\ead{c.papanicolas@cyi.ac.cy}



\affiliation[3]{organization={
The Cyprus Academy of Sciences, Letters and Arts},
            country={Cyprus}}

\cortext[1]{Corresponding author}

\fntext[1]{}


\begin{abstract}
We present an experimental setup and methodology designed to facilitate high-precision thermal measurements required for infrared medical tomography. The approach which is best suited for the study of specialized hardware phantoms comprises  a controlled environmental enclosure, infrared detection, internal thermal reference elements, and a comprehensive data acquisition counting chain and protocol. Temporal and spatial corrections applied to sequential thermal images and panoramic projections reduce measurement fluctuations resulting in measurement uncertainty to approximately 25~mK.  The capability to resolve weak surface temperature variations, well below 0.1~K, meets the requirement of medical imaging sensitivity. The methodology was validated using wax phantoms with elevated-temperature sources ($\Delta T$ =  1.5  to 10~K). Reconstructed 3D thermal tomographic images of hot spots embedded in hardware phantoms are found to be in quantitative agreement with thermocouple measurements and $\mu CT$ derived source positions. The results demonstrate that the proposed setup and methodology enable high-precision thermal measurements and establish the feasibility of detecting surface temperature variations below 0.1 K, consistent with low-temperature localized internal contrasts ($\Delta T$ = 1–3 K) at subsurface depths of a few centimeters, relevant to biological tissue.
\end{abstract}





\maketitle

\section{Introduction}\label{sec:intro}
Infrared (IR) thermography has been employed since the mid-twentieth century for mapping surface temperature distributions in industrial inspection~\cite{OsornioRios2019, Kim2023, DAccardi2025, OswaldTranta2025, Qu2020, Li2025} and physiological studies~\cite{Liu2025, Kesztyus2023,  Ring2007, Lawson1963, Kennedy2009}. Early medical investigations generated considerable interest due to its non-invasive and non-ionizing character; however, clinical adoption remained limited, partly due to inconsistent reproducibility and sensitivity~\cite{Liu2025, Kennedy2009}. Much of this early work preceded the availability of modern infrared detectors, digitized multi-frame acquisition, and controlled measurement environments. Contemporary instrumentation now enables millikelvin (mK) level stability and systematic calibration procedures that were previously unattainable, motivating renewed and more rigorous examination of infrared-based diagnostic methodologies.

Infrared tomography extends planar thermography by attempting to infer internal temperature distributions from surface measurements acquired at multiple viewing angles~\cite{Papanicolas2023, Ledwon2022, Koutsantonis2019, Sage2021, Leontiou2024}. Two physically distinct approaches can be identified, each characterized by a fundamentally different forward model. In \textbf{dynamic infrared tomography}, as explored for example by Koutsantonis et al.~\cite{Koutsantonis2019}, internal heat sources are detected directly through the infrared radiation they emit, exploiting partial transmissivity of the medium in the long-wave infrared (8–14 $\mu m$) spectral region. Reconstruction in this case is governed by radiative transfer and depends on the detectability of photons originating from the sources themselves. Such approaches require relatively large temperature contrasts and favorable propagation conditions in the intervening medium. Consequently, dynamic infrared tomography is generally ill-suited for physiological conditions, where internal local temperature elevations are typically modest (1–3~K) and biological tissue is largely opaque in the long-wave infrared band.

\textbf{Steady-state infrared tomography} offers an alternative approach more compatible with medical applications. In this regime, internal thermal sources are not directly observable radiatively. Instead, heat generated at depth reaches the surface through diffusive transport governed by the stationary heat equation. The measurable quantity is therefore a small boundary temperature perturbation induced by internal thermal heterogeneity. Localized internal temperature elevations of 1–3~K at depths of a few centimeters induce surface variations well below 0.1~K. The resulting inverse problem is intrinsically ill-posed, with reconstruction critically dependent on millikelvin-level measurement stability and accurate characterization of the forward model.

Extracting reliable internal information under steady-state conditions constitutes a problem involving three interconnected components. First, millikelvin-level thermal metrology and environmental stability are required to ensure that measured boundary variations reflect the physical distribution of internal heat sources rather than instrumental drift or ambient fluctuations. Second, the reconstruction methods inevitably depends on material parameters—including thermal conductivity, convective heat-transfer coefficients, emissivity, and possible structural heterogeneity—which may be imperfectly known and may require in-situ characterization or parameterization. Third, robust inverse reconstruction methodologies—such as AMIAS/ RISE \cite{Papanicolas2012, Papanicolas2012arxiv, Alexandrou2015, 9060020, 9875921} —are required to address the ill-conditioned diffusion problem. Reliable steady-state infrared tomography becomes feasible only when these three elements operate coherently. The methodological framework adopted here draws on concepts developed in emission tomography, particularly in SPECT, and on the RISE methodology, with emphasis on system calibration, low-signal reconstruction, and uncertainty quantification~\cite{tomography12040049}.

Recent advances in physics-guided and data-driven reconstruction methodologies, including AI-assisted approaches, have significantly expanded the ability to extract weak signals from boundary measurements. However, as reconstruction methodologies evolve, the need for rigorously controlled and quantitatively characterized experimental infrastructures becomes increasingly important.

While substantial effort has been devoted to reconstruction algorithms~\cite{Papanicolas2012, Kumari2024, Adler2017, Jiang2025, Zeng2022, Leontiou2024, Papanicolas2023}, comparatively little attention has been given to the metrological and environmental conditions required to acquire projection data compatible with steady-state infrared tomography in the physiological temperature regime. In practice, the achievable sensitivity of infrared tomographic measurements is determined not solely by detector specifications but by the stability and calibration of the entire measurement chain, which must be characterized experimentally through a controlled acquisition protocol.

In this work we present the design, implementation, and performance characterization of an instrumentation laboratory developed specifically for steady-state medical infrared tomography. We describe the environmental stabilization strategy, calibration architecture, data-acquisition protocol, and temporal and spatial correction procedures -that are widely applicable and they are not limited to the specific thermal camera model we used- and we quantify the achieved measurement uncertainty. By establishing this controlled metrological foundation, the present work enables systematic investigation of low-contrast thermal tomography under physiologically relevant conditions.
\section{Design Requirements for an Infrared Imaging Laboratory}\label{sec:design}
As discussed in Section~\ref{sec:intro}, the inconsistent results historically reported in medical thermography are likely associated with insufficient environmental control, calibration stability, and unquantified instrumental drift. Infrared tomography places substantially tighter demands on measurement stability because it combines multiple projections acquired over extended acquisition intervals. Small temporal or spatial instabilities can therefore propagate systematically into re-constructed images. The laboratory must  be designed so that instrumental and environmental fluctuations remain well below the smallest diagnostically relevant boundary perturbations.

\subsection{Detectability Requirements}\label{subsec:detect}
The performance targets of the laboratory are  determined by the smallest surface temperature perturbations that must be detected reliably. In steady-state infrared tomography, internal thermal sources are observable only through boundary perturbations resulting from diffusive heat transport. The magnitude of these perturbations depends primarily on source temperature and depth and the thermal properties of the intervening tissue. For physiologically relevant internal temperature elevations $(\Delta T \approx 1$--$3\ \mathrm{K})$, sources located a few centimeters below the surface may produce boundary variations well below 0.1 K.

Reconstruction of internal temperature distributions from boundary measurements constitutes a separate methodological problem and the treatment of the resulting ill-conditioned inverse problem. Several reconstruction approaches have been proposed, including probabilistic methods and physics-guided inversion frameworks~\cite{Papanicolas2023, Papanicolas2012arxiv, Alexandrou2015}. A detailed discussion lies beyond the scope of this paper; the present work focuses on the experimental and metrological foundations required to obtain projection data suitable for such reconstruction.

Reliable detection therefore requires measurement stability well below the 0.1~K level. In addition to detector–counting chain limitations, environmental and boundary-condition fluctuations—such as small air currents affecting convective heat transfer, slow ambient temperature drift, or physiological motion—can produce temperature variations comparable to the signal itself. The combined contributions of instrumental and environmental effects must therefore remain significantly below the expected surface signal, leading to a measurement capability on the order of 25–50~mK. The selection of this range reflects a system-level design criterion informed by detailed modeling and experimental studies, representing a practical operating point that balances achievable performance with constraints imposed by environmental stability and detector characteristics. More stringent targets, although achievable at considerable cost, would likely be limited by uncontrolled boundary-condition variability, particularly in studies involving human subjects. Achieving this level of performance represents a substantial advance in measurement capability within infrared thermography and marks a significant step toward demonstrating the practical feasibility of infrared tomography for medical applications.

\subsection{Thermal Metrology Requirements}\label{subsec:therm_metro}
In medical infrared imaging, diagnostically relevant information derives from spatial temperature differences rather than absolute temperature. Consequently, the laboratory need not determine absolute temperature with high accuracy but must ensure high relative stability and spatial consistency. Calibration can therefore rely on internal referencing.

To enable quantitative tomographic reconstruction, the system must achieve at minimum:

\begin{itemize}
\item Relative temperature resolution $\leq 0.025$ K
\item Operational precision (including environmental contributions) $\leq 0.05$ K
\item Pixel-to-pixel spatial non-uniformity (after correction) $\leq 0.03$ K
\item Frame-to-frame drift during a full acquisition cycle $\leq 0.03$ K
\end{itemize}
Here “full acquisition cycle” refers to the temporal stability maintained over practical tomographic acquisition intervals of up to approximately 30 minutes. Because projections are acquired sequentially, systematic drift during this period would introduce artificial angular structure in the sinogram and propagate into the reconstructed volume.

The 0.05 K precision target ensures that residual instrumental variations remain significantly smaller than medically relevant surface perturbations.

\subsection{Environmental Control Requirements}
Environmental stability is critical because the phantom or subject continuously exchanges heat with its surroundings. Variations in ambient temperature or airflow modify boundary conditions of the heat equation and can introduce perturbations comparable to the diagnostic signal. Detector specifications alone do not determine system performance, which is ultimately limited by calibration stability and environmental control.

The laboratory must therefore ensure that:
\begin{itemize}
\item Temperature variations inside the measurement enclosure during acquisition remain $\leq 0.25$~K (preferably $\leq 0.1$~K)
\item Airflow near the object is suppressed to avoid convective cooling gradients
\item Heat-generating electronics are thermally isolated from the measurement volume
\end{itemize}

Metallic components within the camera field of view should be avoided. If unavoidable, they must be coated with high-emissivity material or shielded to suppress reflective artifacts and radiative coupling.

Stable boundary conditions are therefore an intrinsic requirement of the measurement system rather than a secondary refinement.

\subsection{Measurement and Detection Requirements}\label{subsec:measur_det}
The infrared detector must operate within a noise regime compatible with the stability targets defined above. Although the Noise Equivalent Temperature Difference (NETD) does not alone determine final accuracy, it defines the baseline achievable sensitivity.

Suitable detector characteristics include:

\begin{itemize}
\item NETD $\leq 25$~mK
\item Temporal stability over 30 minutes $\leq 50$~mK
\item Spatial non-uniformity after correction $\leq 30$~mK
\end{itemize}

Sequential acquisition requires internal thermal references. A small number of emitters with highly stable temperature should remain within the camera field of view and be monitored independently. These references provide in-frame calibration anchors for correcting additive drift and slow instrumental variation. Their temperature stability should remain within approximately below 20~mK during acquisition, and emissivity should exceed 0.95 to minimize reflection.

Because tomographic reconstruction combines multiple projections statistically, small systematic biases can accumulate. Reference-based correction is therefore structurally required.

\subsection{Mechanical Precision and Reproducibility}
Tomographic reconstruction requires precise and reproducible angular sampling. 
Mechanical performance targets include:

\begin{itemize}
\item Angular step reproducibility $\leq 0.02^\circ$
\item Centering accuracy relative to the rotation axis $\leq 0.5$~mm
\item Mechanical wobble $\leq 0.2$~mm
\end{itemize}

Rotation must be sufficiently slow to avoid airflow-induced cooling or transient thermal disturbances.

When imaging compliant objects—such as gels, biological tissue, small animals, or human subjects—additional motion and deformation may occur between projections. 
The system should therefore allow:

\begin{itemize}
\item Placement of thermal fiducial markers visible in infrared images
\item Definition of a stable geometric reference frame
\item Registration and correction of relative motion or deformation
\item Verification of positional consistency across repeated acquisitions
\end{itemize}

Reference markers can serve both as thermal calibration anchors and geometric registration points, which becomes essential when transitioning from rigid phantoms to biological structures.

\subsection{Phantom Requirements}\label{subsec:ph_req}
It is desirable that validation phantoms reproduce key thermal properties of biological tissue. 
In particular they should:

\begin{itemize}
\item Possess thermal conductivity in the range $0.2$--$0.6$~W/(mK)~\cite{ElBrawany2009, RodriguezDeRivera2022}
\item Allow embedding of controlled heat sources ($\Delta T$ up to $\sim 10$~K for calibration studies)
\item Incorporate independent internal temperature sensors
\item Provide geometrically characterized internal structure with positional uncertainty $\leq 1$--$2$~mm
\end{itemize}

Such characteristics enable quantitative evaluation of reconstruction accuracy and forward-model fidelity.

\section{Implementation of the Test \& Development Laboratory}\label{sec:implem}
As part of a research program in quantitative steady-state medical infrared tomography, a dedicated test and development laboratory was implemented to address the performance requirements defined in Section~\ref{sec:design} within practical experimental constraints. The system integrates environmental stabilization, geometric control, calibrated thermal referencing, and automated acquisition, with particular emphasis on enabling reliable temporal and spatial correction of projection data prior to inversion. 
\subsection{Environmental Stabilization Architecture}\label{subsec:env_stab}
\begin{figure*}
	\includegraphics[width=1.0\textwidth]{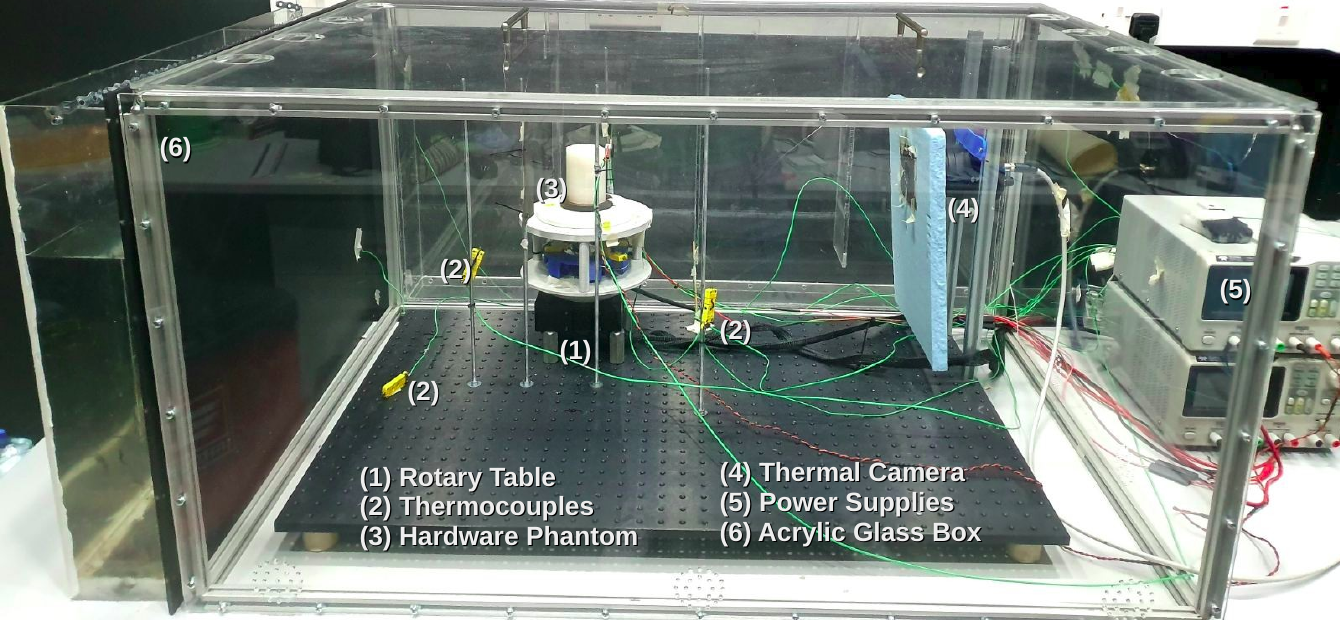}
	\caption{Experimental setup for the study of hardware phantoms  in infrared tomography. The acrylic enclosure stabilizes the local thermal environment by suppressing convective air currents and reducing radiative exchange with the surrounding laboratory. Heat-generating electronics are located outside the enclosure; a water-filled container, the left wall of the enclosure opposite to the camera provides a radiatively uniform cold background behind the phantom improving contrast  and reducing reflective artifacts.}
	\label{fig:setup}
\end{figure*}

The measurement assembly was housed in a temperature-controlled laboratory and enclosed within a custom acrylic-glass housing (Figure~\ref{fig:setup}) to suppress convective fluctuations and isolate the phantom from room-scale thermal drift. The enclosure serves not only as a thermal buffer but also as a means to suppress convective heat transfer and stabilize radiative boundary conditions. By minimizing air exchange and isolating the measurement volume, it reduces temporal fluctuations that would otherwise directly perturb the surface temperature field.

Ventilation openings were incorporated to prevent heat generated by electronic components (thermal camera and rotary table motor) from accumulating inside the enclosure. All major heat-generating equipment—including power supplies, data acquisition units, and the control computer—was therefore positioned outside the enclosure to avoid unintended thermal loading of the measurement volume.

Because steady-state infrared tomography is sensitive to small variations in boundary conditions, the enclosure serves not only as mechanical shielding but also as a means of stabilizing the convective and radiative heat exchange at the phantom surface. Maintaining such boundary stability is essential for ensuring that observed surface temperature variations reflect internal thermal structure rather than environmental perturbations.

A water-filled container was placed behind the phantom, facing the camera, to provide a radiatively uniform background and improve contrast stability during acquisition.

\subsection{Thermal Detection and Referencing Strategy}\label{subsec:det_an_ref}
Thermal imaging was performed using a FLIR A400 infrared camera (24$^\circ$ lens, NETD $< 40$~mK; accuracy $\pm 2$~K)~\cite{FLIRA400Manual}, with emissivity set at 0.95. While absolute accuracy is limited and the NETD performance noisier than desired, the required high relative stability and reproducibility are achieved through internal referencing.

Two ceramic resistors maintained at controlled temperature slightly above that of the phantom, were positioned within the camera’s field of view. Each resistor was monitored by an embedded thermocouple and coupled to a black-coated metallic surface to ensure high emissivity and spatially uniform temperature. These elements serve as stable temporal reference anchors for frame-to-frame alignment.

\begin{figure}
	\includegraphics[width=1.0\columnwidth]{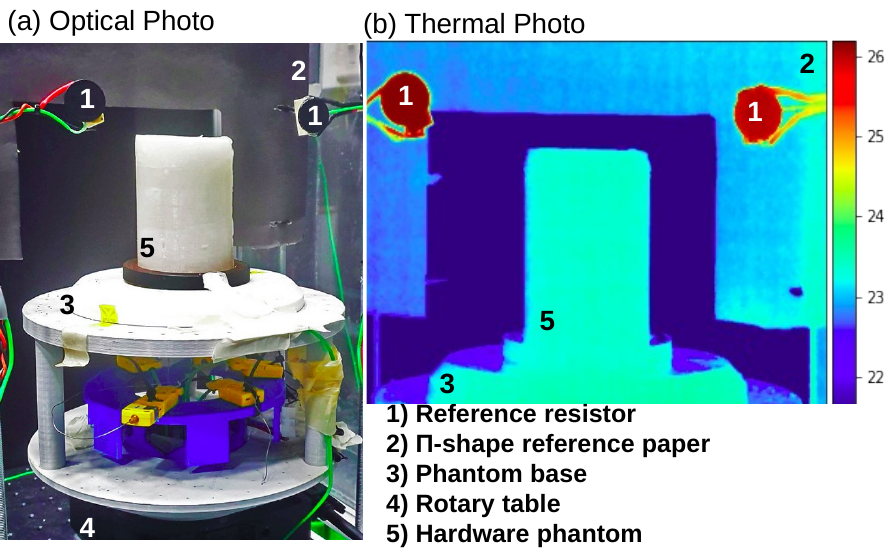}
	\caption{Optical and thermal images of the phantom and reference elements, including temperature-controlled resistors and the $\Pi$-shaped surface. The temperature scale shown is in Degrees °C. The resistors provide stable references for temporal drift correction, while the $\Pi$-surface enables spatial non-uniformity correction, allowing independent temporal and spatial calibration.  The thermal image is displayed using a fixed temperature scale to allow consistent interpretation of temperature variations. \textit{Note: For clarity, since we deal with medical applications, in certain figures and tables we use the more familiar Celsius scale instead of Kelvin.}}
	\label{fig:ref}
\end{figure}

In addition, a $\Pi$-shaped high-emissivity surface (Figure~\ref{fig:ref}) provides an extended isothermal region that enables characterization of spatial non-uniformities in the detector response. Together, these reference elements allow independent control of temporal and spatial calibration effects. This separation simplifies the correction procedure and reduces the risk that temporal drift and spatial non-uniformity become coupled during data processing.

\subsection{Mechanical Positioning and Geometric Stability}\label{subsed:mechan}
The phantom(s) is mounted on a custom plastic base po-sitioned on a motorized rotary table facing to the camera, at a distance of 47~cm. The assembly ensured precise centering with respect to the rotation axis, mechanical stability during acquisition, thermal separation from the motor, and unrestricted routing of electrical leads and thermocouples without cable twisting.

The rotation speed is kept low to avoid airflow-induced cooling or transient convective disturbances. This configuration ensures geometric reproducibility of projection angles and minimized mechanically induced thermal artifacts.

This design ensures that mechanical and geometric uncertainties remain negligible compared to the metrological limits discussed in Section~\ref{sec:design}. 

\subsection{Data Acquisition and Drift Mitigation}\label{subsec:auto}
Data acquisition is automated using a LabVIEW-controlled system to reduce data taking duration and eliminate operator-induced disturbances. After thermal equilibration, projection images are acquired over 360° with set angular increments. The acquisition sequence is flexible and can be adapted depending on the experimental design and conditions; in this work, projections were acquired in two interleaved sequences ($0^\circ$, $8^\circ$, $16^\circ$, \dots\ followed by $4^\circ$, $12^\circ$, $20^\circ$, \dots) to mitigate systematic bias from slow environmental drift. This strategy distributes temporal drift symmetrically across angular positions and reduces correlation between acquisition time and projection angle.

For each projection, the rotary table positions the phantom and remains stationary during image capture. Thermocouple readings from the phantom interior, enclosure air, and reference resistors are recorded nearly simultaneously with each thermal frame. In the present study, 90 projections were acquired with 4° angular resolution, and the complete acquisition cycle lasted approximately 20 minutes.

\subsection{Calibration and Correction Framework}\label{subsec:cal}
Quantitative inversion requires systematic correction of instrumental and environmental perturbations. The implemented procedure combines thermocouple alignment, temporal drift correction, and spatial normalization to ensure consistency across all projections.

\textbf{Thermocouple Offset Alignment}: Because reconstruction depends on relative temperature differences rather than absolute temperature, calibration focused on offset consistency. All thermocouples were monitored during a stabilization interval and aligned relative to a calibrated reference sensor to ensure mutual agreement.

\textbf{Temporal Alignment}: Frame-to-frame detector drift was corrected using the in-frame resistive references. For each projection, an additive correction was applied so that pixel values at the reference locations matched the corresponding thermocouple readings.

\begin{figure}
	\includegraphics[width=1.0\columnwidth, height=4.5cm]{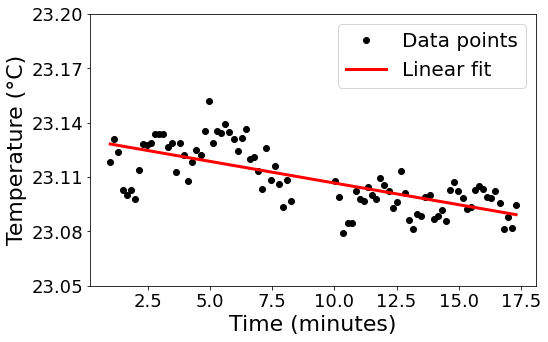}
	\caption{ Air temperature evolution inside the enclosure during a typical acquisition interval ($\sim$20 minutes). The observed linear drift is used to correct slow temporal variations in the projection data.}
	\label{fig:air}
\end{figure}

Slow environmental drift inside the enclosure was modeled using a linear fit to the average air temperature over the acquisition interval (Figure~\ref{fig:air}). Each thermal image was corrected using this drift function, effectively aligning all projections to a common temporal baseline.

\textbf{Spatial Non-Uniformity Correction}: Detector-related spatial gradients were estimated using the $\Pi$-shaped reference surface. Column-wise and row-wise normalization factors were derived from reference region and applied to the entire image. More precisely, for each image column we calculated the ratio between the average value of the upper 10 pixels and the average value of the entire image, and used this ratio to normalize all pixels in that column. Subsequently, each image row was normalized in an analogous manner, using the 10 leftmost pixels as the reference. This process suppresses structured spatial artifacts that would otherwise propagate into the sinograms.

\textbf{Generation of Panoramic Projection Data}
In steady-state infrared tomography, each thermal image records surface temperature rather than a volumetric line integral, as in dynamic infrared tomography. The temperature of a given surface region remains invariant with projection angle, although its apparent position in the camera frame changes.

\begin{figure}
	\includegraphics[width=1.0\columnwidth, height=7cm]{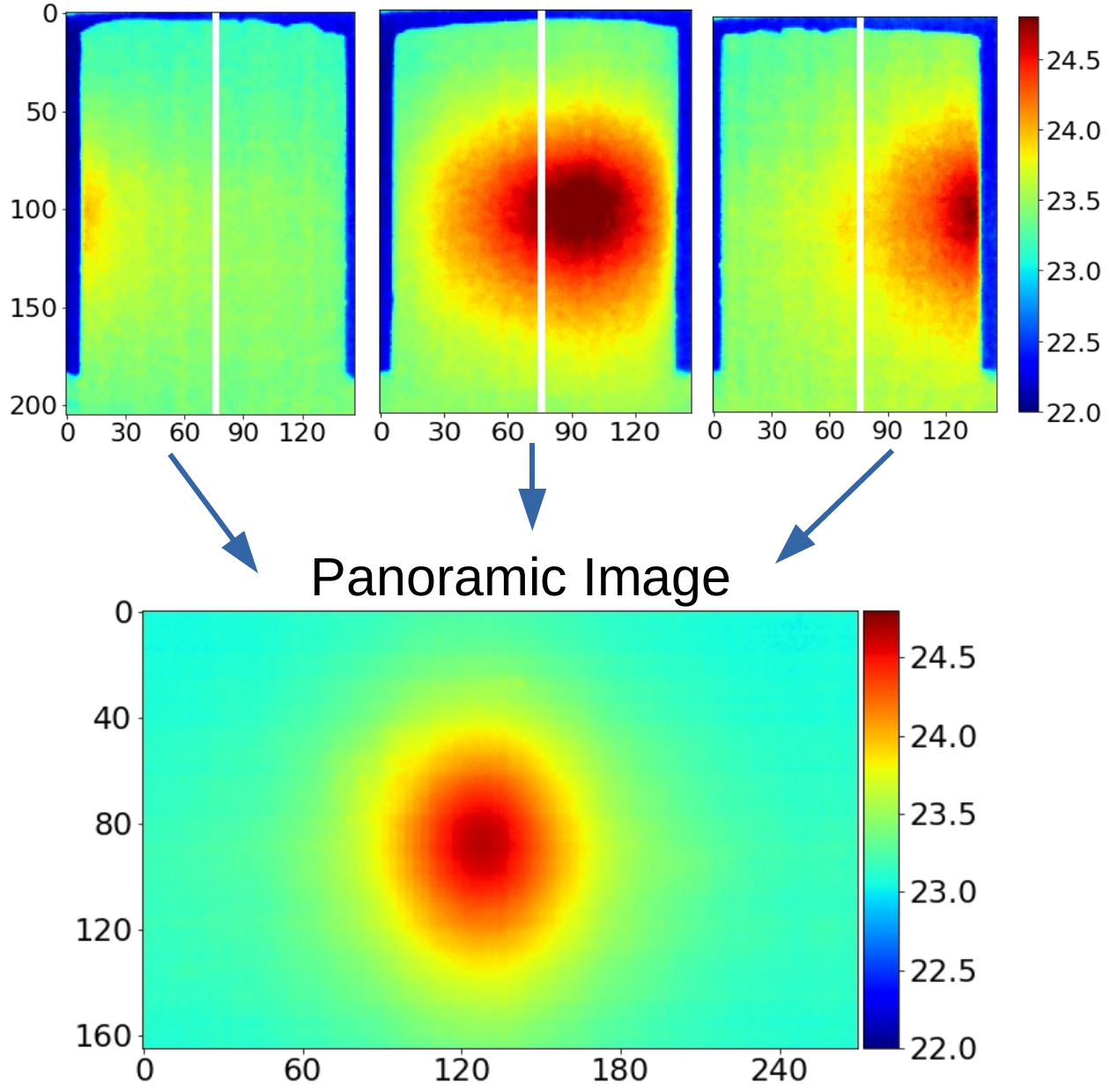}
	\caption{Construction of panoramic projection data from sequential thermal images. At each projection angle, narrow strips corresponding to the front surface of the phantom are extracted and concatenated to form a panoramic dataset. This procedure converts surface temperature measurements into a geometrically consistent representation suitable for tomographic inversion. The thermal images are displayed using a fixed temperature scale.}
	\label{fig:panor}
\end{figure}

Projection data are constructed by extracting narrow vertical strips from the front surface of the phantom at each angle and concatenating them to form a panoramic representation, as illustrated in Figure~\ref{fig:panor}. To reduce stochastic noise, corresponding surface regions are averaged over three consecutive projections. Extracting narrow strips minimizes perspective distortion and ensures that each surface region is analyzed under nearly identical viewing conditions.  This procedure effectively converts a sequence of surface temperature measurements into a geometrically consistent tomographic dataset suitable for inversion.

The resulting corrected panoramic projections constitute the input dataset for the forward model and inversion framework described in the following section. 

Although modern infrared cameras provide high nominal sensitivity, detector specifications alone do not determine the achievable measurement precision in tomographic applications. Environmental boundary fluctuations, detector non-uniformity, and slow temporal drift during extended acquisitions can introduce surface temperature variations comparable to the signal of interest.

The referencing and correction architecture implemented here therefore addresses system-level stability rather than detector sensitivity alone. Moreover, the uncertainty of the measurement chain is not inferred from manufacturer specifications but is determined experimentally through the calibration and acquisition protocol, providing a direct characterization of the effective sensitivity of the tomographic dataset.

\section{Laboratory Performance Characterization}\label{sec:perf}
This section presents the performance achieved by the implemented laboratory and evaluates it against the design requirements defined in Section~\ref{sec:design}. The measurements assess the effectiveness of the adopted system architecture and correction framework in controlling instrumental and environmental perturbations to the level required for quantitative infrared imaging and, in particular, for tomographic reconstruction. The evaluation proceeds hierarchically, examining environmental stability, detector noise, spatial uniformity, projection consistency, and global measurement uncertainty, and concludes with validation of reconstructed temperature distributions against independent internal measurements.

\subsection{Environmental and Temporal Stability}\label{subsec:env_temp}

Temperature stability inside the measurement enclosure was assessed using multiple thermocouples placed within the enclosed volume and in the surrounding laboratory environment. Over extended monitoring periods (24 hours), room temperature fluctuations reached approximately 1 K, whereas variations inside the enclosure remained limited to approximately 0.25 K during acquisition intervals (20-60 minutes). These values satisfy the environmental constraints specified in Section~\ref{sec:design}. Temporal  stability recordings are provided in the Supplementary Material.

\begin{figure}
	\includegraphics[width=1.0\columnwidth]{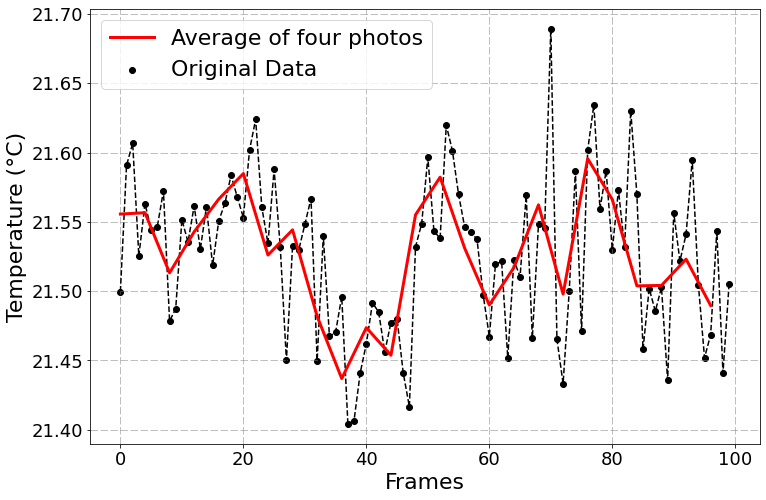}
	\caption{Temporal stability of the thermal image of a uniform phantom. The black curve shows the mean temperature of a central 100-pixel region across successive frames. Frame averaging of four images (red curve) reduces stochastic detector fluctuations from $\sim 300$~mK to $\sim 150$~mK.}
	\label{fig:4ph}
\end{figure}

Short-term detector fluctuations were evaluated by repeatedly imaging a thermally uniform phantom surface under stable conditions. The average temperature of a central region (100 pixels) exhibited frame-to-frame variations up to 300 mK for individual frames (Figure~\ref{fig:4ph}). Averaging four consecutive frames reduced this variation to approximately 150 mK, demonstrating the expected reduction of stochastic detector noise.

This frame averaging was incorporated into the acquisition protocol and served as the first stage of noise suppression prior to temporal correction.

\subsection{Spatial Non-Uniformity and Correction}\label{subsec:spatial}

\begin{figure*}
	\includegraphics[width=1.0\textwidth]{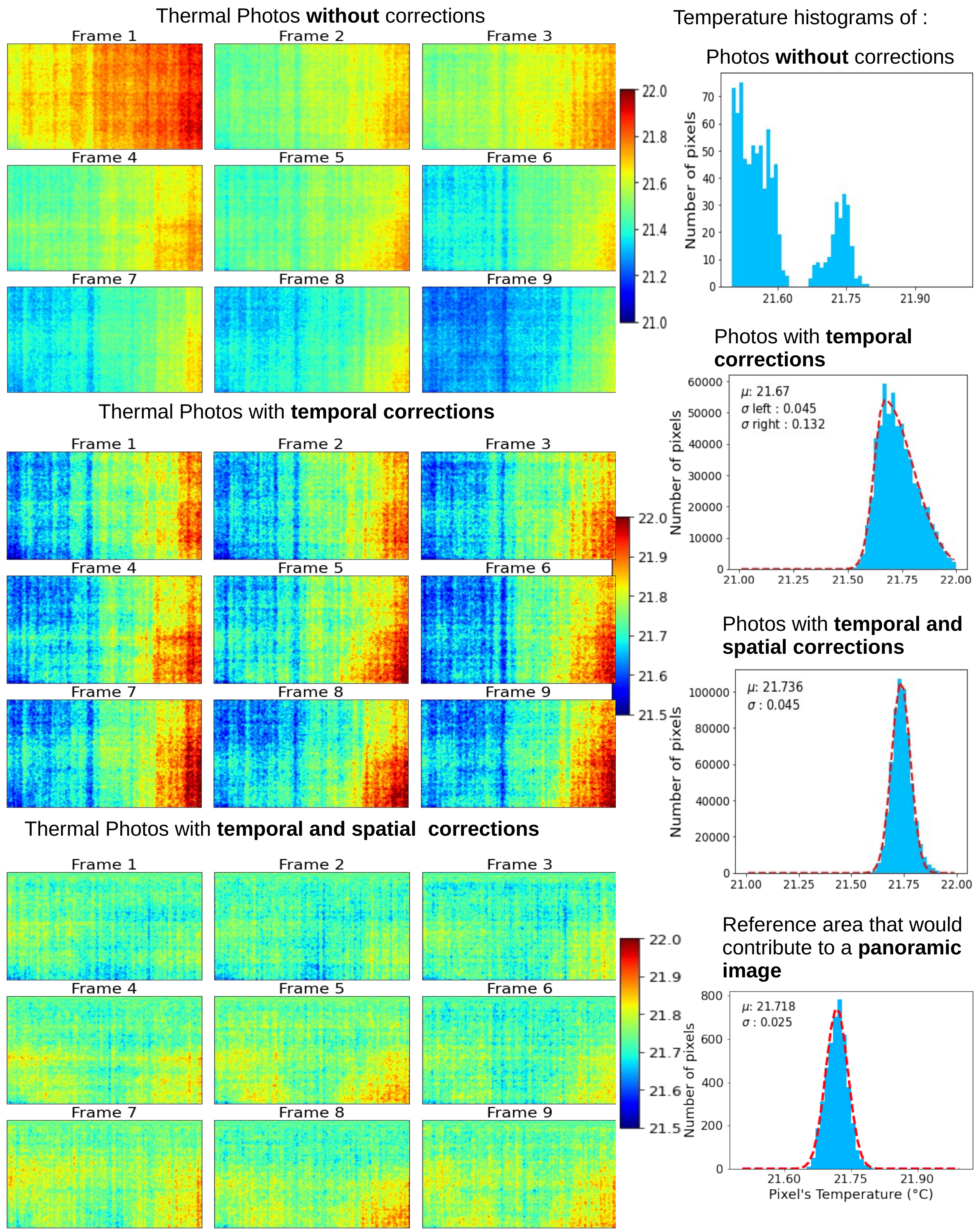}
	\caption{Effect of temporal and spatial correction on a uniform temperature distribution. Top: raw thermal images. Middle: after temporal correction. Bottom: after combined temporal and spatial correction. The right panels show the corresponding pixel temperature histograms, compared with that of the derived panoramic image (Figure~\ref{fig:panor}). After full correction, the temperature distributions become symmetric with $\sigma \approx 45$~mK , while the panoramic image exhibits a reduced spread of approximately $25$~mK.}
	\label{fig:hist_all}
\end{figure*}

Even under uniform environmental thermal conditions, infrared images exhibit pixel-to-pixel variation arising from detector non-uniformity and radiative artifacts. To quantify this effect, a high-emissivity uniform surface (black-painted paper) was imaged before applying the temporal and spatial corrections described in Section~\ref{sec:implem} applied. Figure~\ref{fig:hist_all} presents these images and the corresponding temperature histograms for each of the three cases (uncorrected, temporally corrected, and temporally and spatially corrected). Prior to correction, spatial gradients and column-dependent variations were visible across the field of view. After applying row-wise and column-wise correction factors, the residual  non-uniformity was reduced. The corresponding temperature histograms (Figure~\ref{fig:hist_all}) demonstrate that pixel-to-pixel dispersion was $\leq$ 45~mK.

This correction step is essential, as spatial non-uniformity propagates into structured artifacts in sinograms and degrades inversion stability.

\subsection{Projection Consistency and Sinogram Quality}\label{subsec:quality}

To evaluate the combined impact of temporal and spatial corrections, we examined two-dimensional thermal sinograms acquired from wax phantoms containing embedded resistive heat sources at $\Delta T \approx 1.5$~K. Two configurations were studied: a shallow source (depth 0.8~cm) and a deeper source (depth 3.3~cm).

\begin{figure}
	\includegraphics[width=1.0\columnwidth]{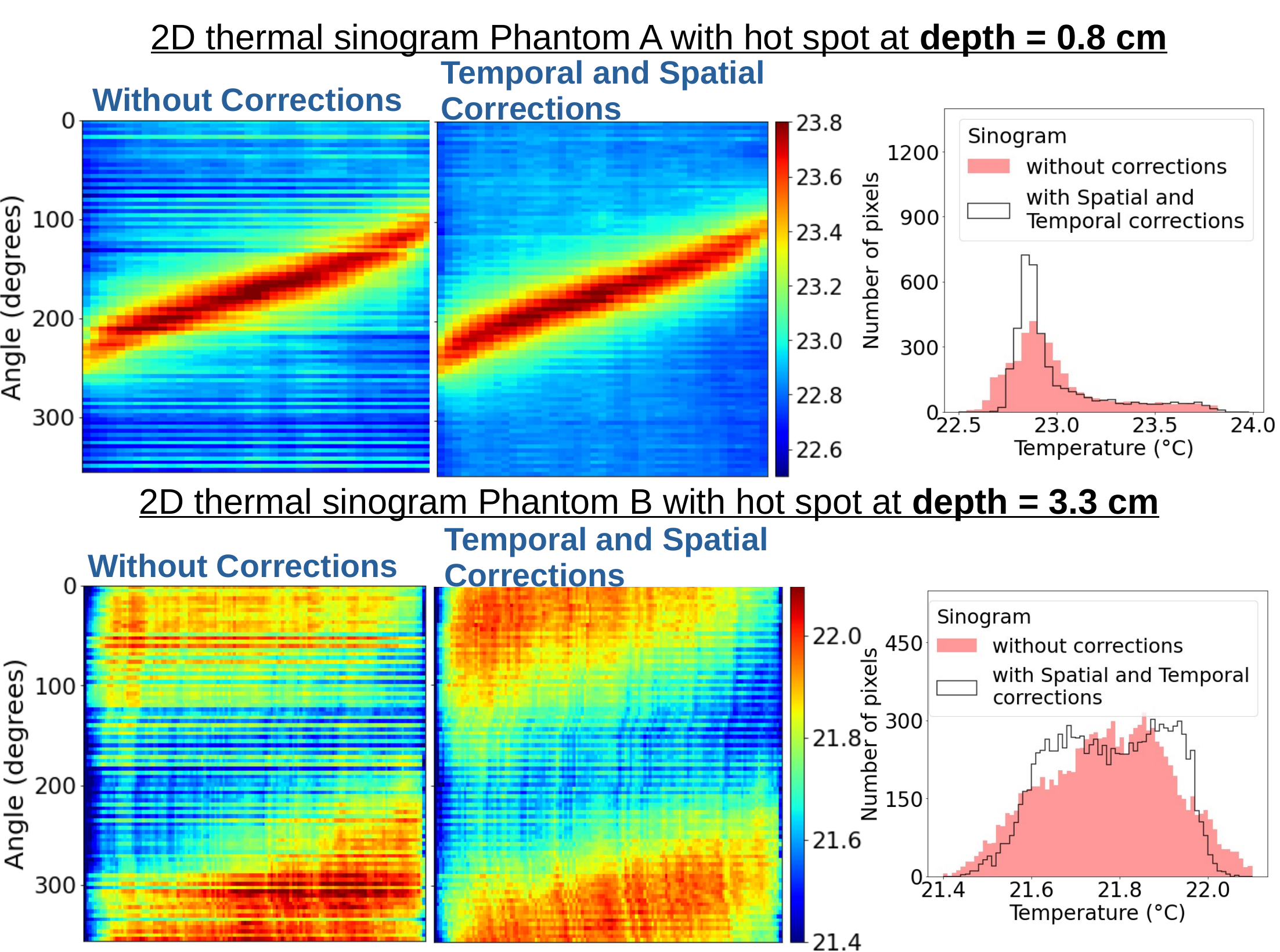}
	\caption{Thermal sinograms and corresponding pixel temperature distributions for wax phantoms containing embedded heat sources at $\Delta T = 1.5~^\circ$C. Left: uncorrected data. Right: after temporal and spatial correction. Top: shallow source (0.8~cm depth). Bottom: deep source (3.3~cm depth). The thermal images are displayed using a fixed temperature scale.}
	\label{fig:hist_sino}
\end{figure}

Figure~\ref{fig:hist_sino} compares sinograms before and after correction. In the uncorrected data, inter-projection temperature offsets and background fluctuations are evident. These variations obscure the underlying signal, particularly for the deeper source where the boundary perturbation is weaker and more diffuse.

After applying temporal alignment and spatial correction, inter-projection variation is substantially reduced. The background temperature distribution narrows, and the signal becomes distinguishable as a structured angular feature. In the deeper-source configuration, the histograms of corrected sinograms exhibits clear separation between background and signal components, where as the uncorrected distribution remains merged and broadened.

These results demonstrate that the implemented correction framework restores angular consistency and yields projection data suitable for quantitative inversion.

\subsection{Global Measurement Uncertainty}\label{subsec:global}
To determine the intrinsic thermal sensitivity of the corrected imaging chain, a uniform high-emissivity surface was imaged using the full acquisition and correction protocol, as presented in Section~\ref{sec:implem}. The combined histogram of all corrected thermal images of reference paper exhibited a Gaussian distribution with standard deviation $\sigma \approx 45$~mK when considering the full frame.

In the final tomographic processing stage, only pixels contributing to the panoramic reconstruction were retained. Restricting analysis to this subset further reduced the dispersion of the Gaussian distribution to $\sigma \approx 25$~mK, as shown in Figure~\ref{fig:hist_all}.

Although modern infrared cameras provide high nominal sensitivity, detector specifications alone do not determine the achievable measurement precision in tomographic applications. Environmental boundary fluctuations, detector non-uniformity, and slow temporal drift during extended acquisitions can introduce surface temperature variations comparable to the signal of interest.

The referencing and correction architecture implemented here therefore addresses system-level stability rather than detector sensitivity alone. Moreover, the uncertainty of the measurement chain is not inferred from manufacturer specifications but is determined experimentally through the calibration and acquisition protocol, providing a direct characterization of the effective sensitivity of the tomographic dataset

The achieved $\sigma \approx 25$~mK uncertainty constitutes the effective sensitivity limit of the laboratory for tomographic projection data. This value lies well below the expected boundary perturbations induced by internal temperature elevations of 1–3 K, thereby meeting the metrological requirements defined in Section~\ref{sec:design}. Further improvements in detector performance would enhance measurement precision and reduce the magnitude of required corrections; however, such improvements are not expected to translate into proportional gains in effective measurement sensitivity in human studies, where system-level effects—such as environmental stability and physiological variability—become dominant.

Table~\ref{tab:uncertainty_budget} summarizes the main sources of uncertainty we had in the data acquisition and the effect of the applied mitigation strategy. The dominant contribution is instrumental, arising primarily from thermal camera fluctuations.

\begin{table}[h]
\centering
\caption{Uncertainty and impact of mitigation strategies in the thermal measurement setup.}
\footnotesize
\begin{tabular}{p{1.3cm} c p{1.0cm} c}
\hline
\textbf{Source of Uncertainty} & \textbf{Typical Magnitude} & \textbf{Action} & \textbf{Mitigation To} \\
\hline
\makecell[l]{Ambient room\\variation} & $\sim$1 K & \makecell{Laboratory\\enclosure} & $\sim$100 mK \\
Detector drift & $\sim$300 mK & Frame averaging & $\sim$150 mK \\
\makecell[l]{Spatial\\non-uniformity} & $\sim$130 mK & Spatial corrections & $\sim$45 mK \\
\hline
\makecell[l]{Effective measurement\\uncertainty} &  & Combined corrections & $\approx$25 mK \\
\hline
\end{tabular}
\label{tab:uncertainty_budget}
\end{table}

\subsection{System Performance Relative to Design Requirements}\label{subsec:syst_perf}

The measurements presented above demonstrate that the implemented laboratory meets, to a substantial extent, the quantitative design requirements defined in Section~\ref{sec:design}. Environmental stability inside the enclosure remained within approximately 0.25~K during acquisition intervals, meeting the boundary-condition stability requirement. 

After temporal alignment and spatial correction, detector-related spatial non-uniformity was reduced to $\leq 45$~mK, while the effective uncertainty of panoramic projection data was approximately 25~mK. These values lie within the stability range required to resolve boundary perturbations associated with physiologically relevant internal temperature elevations. The corrected projection dataset therefore provides a basis compatible with the metrological conditions required for quantitative infrared tomographic analysis.

\subsection{Reconstruction Validation Against Embedded Sensors}\label{subsec:sensors}

A convincing validation of the laboratory design, measurement protocol, and calibration methodology is provided by demonstrating that the system enables measurements relevant to infrared medical imaging, yielding reconstructed internal temperatures in agreement with independent thermocouple measurements embedded within the phantom

Experiments were performed using wax phantoms containing resistive heat sources operated at $\Delta T = 10~\mathrm{K}$. This elevated contrast was selected to validate the forward model and calibration parameters under conditions where parameter uncertainty has minimal relative impact. Under these conditions, the reconstruction yields well-defined temperature distributions, allowing a direct and reliable comparison with independent thermocouple measurements embedded within the phantom.

Three configurations were examined—inner source active, outer source active, and both sources active—providing a range of spatial distributions for validation.  In addition, measurements were performed at lower temperature differences, down to $\Delta T = 1.5~\mathrm{K}$, to assess performance under conditions more representative of medical infrared imaging~\cite{Papanicolas2023}.

\begin{figure*}
	\includegraphics[width=1.0\textwidth]{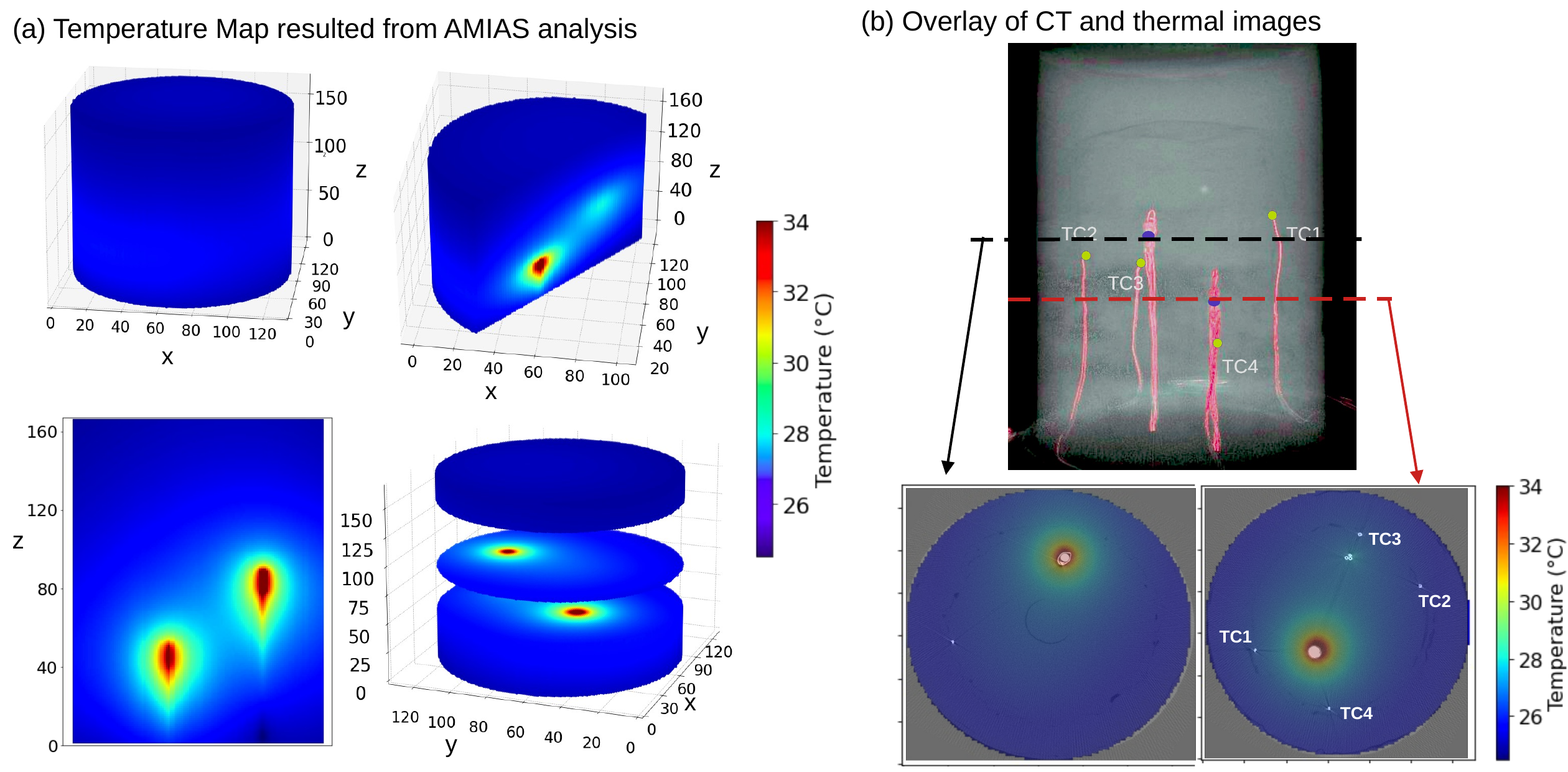}
	\caption{ (a) Three-dimensional temperature distribution reconstructed using the AMIAS /RISE inversion framework from projection data with two embedded resistive sources ($\Delta T = 10~^\circ$C). (b) CT image of the phantom~\cite{biomera2026} with overlaid 2D slices at the resistor level from CT and infrared tomography. The positions of the resistors and thermocouple cables are indicated by white dots demonstrating agreement between CT and thermal tomography.}
	\label{fig:ct1}
\end{figure*}

Projection data were analyzed using the AMIAS/ RISE-based inversion framework, treating source position, power, and ambient temperature as free parameters. The reconstructed three-dimensional temperature distribution is shown in Figure~\ref{fig:ct1}a. The overlaid 2D thermal and CT images (Figure~\ref{fig:ct1}b) show the level of agreement in the predicted positions of the resistors in the $xy$-plane.

Table~\ref{tab:thermocouples} summarizes the comparison between reconstructed temperatures and thermocouple measurements at multiple internal locations. Across all configurations, reconstructed values show reasonable agreement with direct measurements. The remaining discrepancy is likely attributable to limitations of the reconstruction methodology and/or uncertainties in the physical parameters of the thermal diffusion model. These aspects are beyond the scope of the present work and will be addressed in a future publication, with further details provided in the doctoral dissertation of A. Frixou~\cite{frixou2026thermal} which evaluates the performance of the aforementioned setup over a range of $\Delta T$ scenarios.

\begin{table}[ht]
\centering
\small
\caption{Temperature values measured by the four thermocouples, shown in $\mu$CT sections in Figure~\ref{fig:ct1}, and corresponding AMIAS /RISE results; hot-spots at $\Delta T = 10~^\circ$C.}
\label{tab:thermocouples}
\resizebox{\columnwidth}{!}{
\begin{tabular}{lll}
\toprule
Thermocouple & Experimental Value ($^\circ$C) & AMIAS /RISE Results ($^\circ$C) \\
\midrule
\multicolumn{3}{c}{\textbf{Inner resistor on}} \\
\midrule
TC 1 & $24.7 \pm 0.1$ & $24.71 \pm 0.09$ \\
TC 2 & $24.4 \pm 0.1$ & $24.56 \pm 0.07$ \\
TC 3 & $24.2 \pm 0.1$ & $24.56 \pm 0.07$ \\
TC 4 & $25.4 \pm 0.1$ & $25.56 \pm 0.15$ \\
\midrule
\multicolumn{3}{c}{\textbf{Outer resistor on}} \\
\midrule
TC 1 & $23.6 \pm 0.1$ & $23.49 \pm 0.06$ \\
TC 2 & $24.2 \pm 0.1$ & $24.18 \pm 0.15$ \\
TC 3 & $25.7 \pm 0.1$ & $26.00 \pm 0.40$ \\
TC 4 & $23.2 \pm 0.1$ & $23.26 \pm 0.03$ \\
\midrule
\multicolumn{3}{c}{\textbf{Both resistors on}} \\
\midrule
TC 1 & $25.7 \pm 0.1$ & $25.72 \pm 0.09$ \\
TC 2 & $26.1 \pm 0.1$ & $26.27 \pm 0.07$ \\
TC 3 & $27.7 \pm 0.1$ & $28.11 \pm 0.07$ \\
TC 4 & $26.2 \pm 0.1$ & $26.41 \pm 0.15$ \\
\bottomrule
\end{tabular}
}
\end{table}

This level of agreement confirms that the combined system—environmental stabilization, referencing, temporal correction, spatial correction, calibrated forward model, and reconstruction algorithm—operates coherently and yields quantitatively accurate internal temperature estimates.

Collectively, these measurements characterize the achieved system performance and demonstrate that the implemented laboratory design, measurement protocol, and calibration methodology satisfy the metrological requirements defined in Section~\ref{sec:design}. The combined stabilization, referencing, and correction architecture reduces instrumental and environmental perturbations to a level compatible with the detection of weak boundary temperature signals. The resulting projection data therefore provide a reliable basis for quantitative infrared tomographic reconstruction.

\section{Summary and Conclusions}\label{sec:summary}

We have presented the design, implementation, and validation of a dedicated laboratory for steady-state infrared tomography under physiologically relevant thermal conditions. The system integrates environmental stabilization, calibrated thermal referencing, geometric control, and automated acquisition to achieve millikelvin-level measurement stability. The current implementation is designed for controlled hardware phantoms and is not intended for direct studies on human subjects.

Experimental characterization demonstrates that the implemented correction architecture reduces instrumental and environmental perturbations to an effective uncertainty of approximately 25 mK in panoramic projection data. This level of performance is sufficient to resolve boundary temperature variations associated with physiologically relevant internal thermal contrasts at depths of a few centimeters.

A steady-state forward model based on diffusive heat transport was coupled to an AMIAS /RISE-based inversion framework, permitting probabilistic reconstruction of internal heat sources and quantitative uncertainty estimation. This formulation explicitly distinguishes medical infrared tomography from radiative or transmission-based modalities and emphasizes the role of boundary perturbations governed by conductive heat transfer.

Experimental validation demonstrated that the implemented correction architecture reduces systematic artifacts to a level compatible with low-contrast inversion. The achieved projection uncertainty ($\sim 25~\mathrm{mK}$ in panoramic data) lies below the expected surface perturbations induced by physiologically relevant internal temperature elevations (1–3~K). Reconstruction experiments with embedded resistive sources confirmed reasonable quantitative agreement between inferred and independently measured internal temperatures-improved detector performance could further reduce instrumental uncertainty and simplify correction procedures; however, such improvements are not expected to yield proportional gains in effective measurement sensitivity in studies involving human subjects, where system-level effects—such as environmental variability and physiological processes—become dominant.

Collectively, these results establish that steady-state infrared tomography can be performed within a controlled laboratory environment with sufficient metrological rigor to support quantitative tomographic inversion.

The present system is intended for controlled phantom studies and not for direct application to human subjects, but it establishes the measurement stability and calibration framework required for such investigations, where environmental conditions are less stable and not reproducible. The laboratory provides a platform for systematic phantom studies, enabling the development and validation of measurement protocols and reconstruction approaches under well-defined conditions. Such studies are necessary to acquire the experimental know-how required for the application of infrared tomography to human subjects. Continued experimentation will also support the development of reconstruction methodologies adapted to diffusive heat transport, as approaches from other tomographic modalities require substantial modification.
\subsection*{CRediT authorship contribution statement}
\textbf{A. Frixou}: Experimental setup, data curation and analysis, software, writing – original draft. \textbf{E. Stiliaris}: Experimental setup, data curation, writing – review and editing. \textbf{C. N. Papanicolas}: Conceptualization, experimental design,  writing – original draft.
\subsection*{Data Availability}

The data that support this study are available upon reasonable request.

\subsection*{Acknowledgments}
The bulk of the results presented in this work form part of the doctoral dissertation of Anna Frixou at the The Cyprus Institute. The authors gratefully acknowledge the support of the Graduate School of the The Cyprus Institute.  The research activities and the engagement of C.N. Papanicolas were partly supported by the Cyprus Academy of Sciences, Letters and Arts.  The authors would also like to thank Prof. Kirsi Lauentz and Dr Anis Fatima for facilitating access to the BIOMERA platform, and providing the $\mu$CT images of the wax phantoms.






\printcredits

\bibliographystyle{elsarticle-num}

\bibliography{cas-refs}

@article{OsornioRios2019,
  author    = {Osornio-Rios, R. Alfredo and Antonino-Daviu, J. A. and Romero-Troncoso, R. de Jesus},
  title     = {Recent Industrial Applications of Infrared Thermography: A Review},
  journal   = {IEEE Transactions on Industrial Informatics},
  volume    = {15},
  number    = {2},
  pages     = {615--625},
  year      = {2019},
  month     = {February},
  doi       = {10.1109/TII.2018.2884738}
}

@article{Kim2023,
  author    = {Kim, H. and Lamichhane, N. and Kim, C. and Shrestha, R.},
  title     = {Innovations in Building Diagnostics and Condition Monitoring: A Comprehensive Review of Infrared Thermography Applications},
  journal   = {Buildings},
  volume    = {13},
  number    = {11},
  pages     = {2829},
  year      = {2023},
  doi       = {10.3390/buildings13112829}
}

@article{DAccardi2025,
  author    = {D'Accardi, E. and Ammannato, L. and Giannasi, A. and Pieri, M. and Masciopinto, G. and Ancona, F. and Santonicola, G. and Palumbo, D. and Galietti, U.},
  title     = {Infrared Thermography for Non-Destructive Testing of Cooling Hole Integrity and Flow Evaluation in Specimens Made with Innovative Technologies},
  journal   = {Engineering Proceedings},
  volume    = {85},
  pages     = {15},
  year      = {2025},
  doi       = {10.3390/engproc2025085015}
}

@article{OswaldTranta2025,
  author    = {Oswald-Tranta, B.},
  title     = {Inductive Thermography -- Review of a Non-Destructive Inspection Technique for Surface Crack Detection},
  journal   = {Quantitative InfraRed Thermography Journal},
  volume    = {22},
  number    = {5},
  pages     = {478--502},
  year      = {2025},
  doi       = {10.1080/17686733.2024.2448049}
}

@article{Qu2020,
  author    = {Qu, Z. and Jiang, P. and Zhang, W.},
  title     = {Development and Application of Infrared Thermography Non-Destructive Testing Techniques},
  journal   = {Sensors},
  volume    = {20},
  number    = {14},
  pages     = {3851},
  year      = {2020},
  doi       = {10.3390/s20143851}
}

@article{Li2025,
  author    = {Li, Rongcheng and Wang, Fei and Yin, Peng and Yang, Feng and Zhao, Jianghao and Yue, Zhuoyan and Liu, Lixia and Sfarra, Stefano and Vesala, G. T. and Yue, Honghao and Liu, Junyan},
  title     = {A Review of Ultrasonic Infrared Thermography in Non-Destructive Testing and Evaluation (NDT\&E): Physical Principles, Theory, and Data Processing},
  journal   = {Infrared Physics \& Technology},
  volume    = {150},
  pages     = {105961},
  year      = {2025},
  issn      = {1350-4495},
  doi       = {10.1016/j.infrared.2025.105961}
}

@article{Liu2025,
  author  = {Liu, Q. and Li, M. and Wang, W. and others},
  title   = {Infrared Thermography in Clinical Practice: A Literature Review},
  journal = {European Journal of Medical Research},
  volume  = {30},
  pages   = {33},
  year    = {2025},
  doi     = {10.1186/s40001-025-02278-z}
}

@article{Kesztyus2023,
  author  = {Keszty{\"u}s, D. and Brucher, S. and Wilson, C. and Keszty{\"u}s, T.},
  title   = {Use of Infrared Thermography in Medical Diagnosis, Screening, and Disease Monitoring: A Scoping Review},
  journal = {Medicina},
  volume  = {59},
  number  = {12},
  pages   = {2139},
  year    = {2023},
  doi     = {10.3390/medicina59122139}
}

@article{Ring2007,
  author  = {Ring, E. F. J.},
  title   = {The Historical Development of Temperature Measurement in Medicine},
  journal = {Infrared Physics \& Technology},
  volume  = {49},
  number  = {3},
  pages   = {297--301},
  year    = {2007},
  doi = {10.1016/j.infrared.2006.06.029}

}

@article{Lawson1963,
  author  = {Lawson and Chughtai},
  title   = {Breast Cancer and Body Temperature},
  journal = {JAMA: The Journal of the American Medical Association},
  volume  = {183},
  number  = {7},
  pages   = {221},
  year    = {1963},
  url = {https://pmc.ncbi.nlm.nih.gov/articles/PMC1920940/}
}

@article{Kennedy2009,
  author  = {Deborah A. Kennedy and Tanya Lee and Dugald Seely},
  title   = {A Comparative Review of Thermography as a Breast Cancer Screening Technique},
  journal = {Integrative Cancer Therapies},
  volume  = {8},
  number  = {1},
  pages   = {9--16},
  year    = {2009},
  doi = {10.1177/1534735408326171}
}

@inproceedings{Papanicolas2023,
  author    = {C. Papanicolas and A. Frixou and L. Koutsantonis},
  title     = {Thermal Tomography for Medical Applications},
  booktitle = {2023 IEEE Nuclear Science Symposium, Medical Imaging Conference and International Symposium on Room-Temperature Semiconductor Detectors (NSS MIC RTSD)},
  location  = {Vancouver, BC, Canada},
  pages     = {1--1},
  year      = {2023},
  doi       = {10.1109/NSSMICRTSD49126.2023.10337910}
}

@article{Ledwon2022,
  author  = {Ledwon, D. and Sage, A. and Juszczyk, J. and Rudzki, M. and Badura, P.},
  title   = {Tomographic Reconstruction from Planar Thermal Imaging Using Convolutional Neural Network},
  journal = {Scientific Reports},
  volume  = {12},
  pages   = {2347},
  year    = {2022},
  doi ={10.1038/s41598-022-06076-z}
}

@article{Koutsantonis2019,
  author  = {Koutsantonis, L. and Rapsomanikis, A. N. and Stiliaris, E. and Papanicolas, C. N.},
  title   = {Examining an Image Reconstruction Method in Infrared Emission Tomography},
  journal = {Infrared Physics \& Technology},
  volume  = {98},
  pages   = {266--277},
  year    = {2019},
  doi = {10.1016/j.infrared.2019.03.015}
}

@incollection{Sage2021,
  author    = {Sage, A. and Ledwon, D. and Juszczyk, J. and Badura, P.},
  title     = {{3D} Thermal Volume Reconstruction from {2D} Infrared Images—A Preliminary Study},
  booktitle = {Innovations in Biomedical Engineering. Advances in Intelligent Systems and Computing},
  publisher = {Springer},
  address   = {Cham, Switzerland},
  volume    = {1223},
  pages     = {371--379},
  year      = {2021},
  doi = {10.1007/978-3-030-52180-6_38}
}

@article{Leontiou2024,
  author  = {Leontiou, T. and Frixou, A. and Charalambides, M. and Stiliaris, E. and Papanicolas, C. N. and Nikolaidou, S. and Papadakis, A.},
  title   = {Three-Dimensional Thermal Tomography with Physics-Informed Neural Networks},
  journal = {Tomography},
  volume  = {10},
  pages   = {1930--1946},
  year    = {2024},
  doi     = {10.3390/tomography10120140}
}

@misc{Papanicolas2012,
  author    = {Costas N. Papanicolas and Loizos Koutsantonis and Efstathios Stiliaris},
        title={A Novel Analysis Method for Emission Tomography}, 
      author={Costas N. Papanicolas and Loizos Koutsantonis and Efstathios Stiliaris},
      year={2018},
      eprint={1804.03915},
      archivePrefix={arXiv},
      primaryClass={physics.med-ph},
}

@article{Kumari2024,
  author  = {Kumari, S. and Singh, P.},
  title   = {Deep Learning for Unsupervised Domain Adaptation in Medical Imaging: Recent Advancements and Future Perspectives},
  journal = {Computers in Biology and Medicine},
  volume  = {170},
  pages   = {107912},
  year    = {2024},
  doi     = {10.1016/j.compbiomed.2023.107912}
}

@article{Adler2017,
  author  = {Adler, J. and Öktem, O.},
  title   = {Solving Ill-Posed Inverse Problems Using Iterative Deep Neural Networks},
  journal = {Inverse Problems},
  volume  = {33},
  pages   = {124007},
  year    = {2017},
  doi     = {10.1088/1361-6420/aa9581}
}

@article{Jiang2025,
  author  = {Jiang, H. and Zhang, Q. and Hu, Y. and Jin, Y. and Liu, H. and Chen, Z. and Zhao, Y. and Fan, W. and Zheng, H. and Liang, D. and others},
  title   = {Memory-Enhanced and Multi-Domain Learning-Based Deep Unrolling Network for Medical Image Reconstruction},
  journal = {Physics in Medicine \& Biology},
  volume  = {70},
  pages   = {175008},
  year    = {2025},
  doi     = {10.1088/1361-6560/adf939}
}

@article{Zeng2022,
  author  = {Zeng, D. and Zeng, C. and Zeng, Z. and Li, S. and Deng, Z. and Chen, S. and Bian, Z. and Ma, J.},
  title   = {Basis and Current State of Computed Tomography Perfusion Imaging: A Review},
  journal = {Physics in Medicine \& Biology},
  volume  = {67},
  pages   = {18TR01},
  year    = {2022},
  doi     = {10.1088/1361-6560/ac8717}
}

@article{Papanicolas2012arxiv,
  author  = {Papanicolas, C. N. and Stiliaris, E.},
  title   = {A novel method of data analysis for hadronic physics},
  eprint  = {1205.6505},
  archivePrefix = {arXiv},
  year    = {2012}
}

@article{Alexandrou2015,
  author  = {Alexandrou, C. and Leontiou, T. and Papanicolas, C. N. and Stiliaris, E.},
  title   = {Novel Analysis Method for Excited States in Lattice QCD: The Nucleon Case},
  journal = {Physical Review D},
  volume  = {91},
  number  = {1},
  pages   = {014506},
  year    = {2015},
  doi     = {10.1103/PhysRevD.91.014506}
}

@article{ElBrawany2009,
  author  = {El-Brawany, M. A. and Nassiri, D. K. and Terhaar, G. and Shaw, A. and Rivens, I. and Lozhken, K.},
  title   = {Measurement of Thermal and Ultrasonic Properties of Some Biological Tissues},
  journal = {Journal of Medical Engineering \& Technology},
  volume  = {33},
  number  = {3},
  pages   = {249--256},
  year    = {2009},
  pmid    = {19340696},
  doi = {10.1080/03091900802451265}
}

@article{RodriguezDeRivera2022,
  author  = {Rodríguez de Rivera, Pedro Jesús and Rodríguez de Rivera, Miriam and Socorro, Fabiola and Calbet, Jose A. L. and Rodríguez de Rivera, Manuel},
  title   = {Advantages of in Vivo Measurement of Human Skin Thermal Conductance Using a Calorimetric Sensor},
  journal = {Journal of Thermal Analysis and Calorimetry},
  volume  = {147},
  pages   = {10027--10036},
  year    = {2022},
  doi = {10.1007/s10973-022-11275-x}
}

@manual{FLIRA400Manual,
  title        = {{FLIR A400} User Manual},
  organization = {FLIR Systems},
  year         = {n.d.},
  url          = {https://thermokameras.com/Verkauf/Flir%20A-Serie/Manual%20A400%20Smart%20Sensor%20configuration.pdf},
  note         = {Accessed: 2026-03-10}
}

@INPROCEEDINGS{9060020,

  author={Lemesios, Christos and Koutsantonis, Loizos and Papanicolas, Costas N.},

  booktitle={2019 IEEE Nuclear Science Symposium and Medical Imaging Conference (NSS/MIC)}, 

  title={RISE: Tomographic Image Reconstruction in Positron Emission Tomography}, 

  year={2019},

  volume={},

  number={},

  pages={1-4},

  doi={10.1109/NSS/MIC42101.2019.9060020}}

@misc{biomera2026,
  author       = {{BioMERA Platform}},
  title        = {BioMERA: Platform for Biosciences and Human Health in Cyprus},
  year         = {2026},
  url          = {https://biomera.cyi.ac.cy/},
  note         = {Accessed: 2026-04-05}
}

@Article{tomography12040049,
AUTHOR = {Frixou, Anna and Leontiou, Theodoros and Stiliaris, Efstathios and Papanicolas, Costas N.},
TITLE = {Standardized Images and Evaluation Metrics for Tomography},
JOURNAL = {Tomography},
VOLUME = {12},
YEAR = {2026},
NUMBER = {4},
ARTICLE-NUMBER = {49},
ISSN = {2379-139X},
DOI = {10.3390/tomography12040049}
}

@INPROCEEDINGS{9875921,

  author={Keliri, Aikaterini and Koutsantonis, Loizos and Stiliaris, Efstathios and Parpottas, Yiannis and Charitou, Giorgos and Panagi, Sotiris and Papanicolas, Costas N.},

  booktitle={2021 IEEE Nuclear Science Symposium and Medical Imaging Conference (NSS/MIC)}, 

  title={Application of RISE in SPECT Myocardial Perfusion Imaging, using a Cardiac Phantom}, 

  year={2021},

  volume={},

  number={},

  pages={1-5},

  doi={10.1109/NSS/MIC44867.2021.9875921}}

@phdthesis{frixou2026thermal,
  author    = {Anna Frixou},
  title     = {Thermal Tomography for Medical Applications},
  school    = {The Cyprus Institute},
  address   = {Nicosia, Cyprus},
  year      = {2026},
  type      = {PhD thesis}
}



\end{document}